\documentclass[12pt]{article}
  
\catcode`@=11  
\newcount\@tempcntc  
\def\@citex[#1]#2{\if@filesw\immediate\write\@auxout{\string\citation{#2}}\fi  
  \@tempcnta\z@\@tempcntb\m@ne\def\@citea{}\@cite{\@for\@citeb:=#2\do  
    {\@ifundefined  
       {b@\@citeb}{\@citeo\@tempcntb\m@ne\@citea\def\@citea{,}{\bf  
?}\@warning  
       {Citation `\@citeb' on page \thepage \space undefined}}%
    {\setbox\z@\hbox{\global\@tempcntc0\csname b@\@citeb\endcsname\relax}%
     \ifnum\@tempcntc=\z@ \@citeo\@tempcntb\m@ne  
       \@citea\def\@citea{,}\hbox{\csname b@\@citeb\endcsname}%
     \else  
      \advance\@tempcntb\@ne  
      \ifnum\@tempcntb=\@tempcntc  
      \else\advance\@tempcntb\m@ne\@citeo  
      \@tempcnta\@tempcntc\@tempcntb\@tempcntc\fi\fi}}\@citeo}{#1}}  
\def\@citeo{\ifnum\@tempcnta>\@tempcntb\else\@citea\def\@citea{,}%
  \ifnum\@tempcnta=\@tempcntb\the\@tempcnta\else  
   {\advance\@tempcnta\@ne\ifnum\@tempcnta=\@tempcntb \else  
\def\@citea{--}\fi  
    \advance\@tempcnta\m@ne\the\@tempcnta\@citea\the\@tempcntb}\fi\fi}  
\catcode`@=12  
\newcommand{\re}{\mathop{\rm{Re}}\nolimits} 
\newcommand{\im}{\mathop{\rm{Im}}\nolimits} 
 
\begin{document} 
  
\title{\vskip-3cm{\baselineskip12pt  
\centerline{\normalsize \hfill UB-ECM-PF~01/11}  
\centerline{\normalsize \hfill DESY~01-120}  
\centerline{\normalsize \hfill NYU-TH/01/08/09}  
\centerline{\normalsize \hfill hep-ph/0109228}  
\centerline{\normalsize \hfill September~2001}}  
\vskip1.5cm   
Width and Partial Widths of Unstable Particles in the Light of the Nielsen 
Identities}  
\author{{\sc Pietro A. Grassi,$^1$ Bernd A. Kniehl,$^2$\thanks{%
Permanent address: II. Institut f\"ur Theoretische Physik, Universit\"at 
Hamburg, Luruper Chaussee 149, 22761 Hamburg, Germany.} 
Alberto Sirlin$^1$}\\  
{\normalsize $^1$ Department of Physics, New York University,}\\  
{\normalsize 4~Washington Place, New York, NY~10003, USA}\\  
{\normalsize $^2$ Departament d'Estructura i Constituents de la Mat\`eria, 
Universitat de Barcelona,}\\  
{\normalsize Avinguda Diagonal, 647, 08028 Barcelona, Spain}} 
  
\date{}  
  
\maketitle  
\thispagestyle{empty}  
  
\begin{abstract}  
Fundamental properties of unstable particles, including mass, width, and 
partial widths, are examined on the basis of the Nielsen identities (NI) that 
describe the gauge dependence of Green functions. 
In particular, we prove that the pole residues and associated definitions of 
branching ratios and partial widths are gauge independent to all orders. 
A simpler, previously discussed definition of branching ratios and partial 
widths is found to be gauge independent through next-to-next-to-leading order. 
It is then explained how it may be modified in order to extend the gauge 
independence to all orders. 
We also show that the physical scattering amplitude is the most general 
combination of self-energy, vertex, and box contributions that is gauge 
independent for arbitrary $s$, discuss the analytical properties of the NI 
functions, and exhibit explicitly their one-loop expressions in the 
$Z$-$\gamma$ sector of the Standard Model. 
 
\smallskip 
  
\end{abstract}  
  
\newpage  

\section{INTRODUCTION}

The conventional definitions of the mass and width of unstable particles are  
\begin{eqnarray}  
\label{def_mass_1}  
M^2&=&M_0^2+\re A(M^2),\\  
\label{def_width_1}  
M\Gamma&=&-\frac{\im A(M^2)}{1-\re A^\prime(M^2)}, 
\end{eqnarray}  
where $M_0$ is the bare mass and $A(s)$ is the self-energy in the case of 
scalar bosons and the transverse self-energy in the case of vector bosons. 
The partial widths are defined by expressing the numerator of  
Eq.~(\ref{def_width_1}) as a sum of cut contributions involving distinct sets 
of final-state physical particles. 
We will refer to $M$ as the on-shell mass and to Eqs.~(\ref{def_mass_1}) and  
(\ref{def_width_1}) as the conventional on-shell formulation.  
  
However, it was shown in Ref.~\cite{si_1} that, in a gauge theory,  
Eqs.~(\ref{def_mass_1}) and (\ref{def_width_1}) become gauge dependent in  
next-to-next-to-leading order (NNLO), i.e.\ in $O(g^4)$ and $O(g^6)$, 
respectively, where $g$ is a generic gauge coupling. 
In the same papers, it was proposed that a way of solving this problem is to  
base the definitions of mass and width on the complex-valued position of the 
propagator's pole:  
\begin{equation}  
\label{def_mass_2}  
\bar s=M_0^2+A(\bar s).   
\end{equation} 
Employing the parameterization $\bar s=m_2^2-im_2\Gamma_2$ of 
Ref.~\cite{si_1}, we have   
\begin{eqnarray}  
m_2^2&=&M_0^2+\re A(\bar s), 
\label{def_width_2}\\  
m_2\Gamma_2&=&-\im A(\bar s). 
\label{def_width_2.1}  
\end{eqnarray}  
From Eq.~(\ref{def_width_2}), we see that the mass counterterm is given by   
$\re A(\bar s)$, rather than $\re A(M^2)$.  
  
In the recent past, a number of authors have advocated definitions of mass and 
width based on $\bar s$ \cite{zmass}, and the conclusions of Ref.~\cite{si_1}
have been confirmed by later studies \cite{psz,si-kniehl} and proven to all
orders \cite{gg}.  
In particular, it has been shown that the gauge dependences of $M$ and 
$\Gamma$ are numerically large in the case of a heavy Higgs boson 
\cite{si-kniehl}.  
It has also been pointed out that Eq.~(\ref{def_width_1}) leads to serious 
problems if $A(s)$ is not analytic in the neighborhood of $M^2$, a situation 
that occurs when $M$ is very close to a physical threshold \cite{bhatta} or,
in the resonance region, when the unstable particle is coupled to massless 
quanta, as in the cases of the $W$ boson and unstable quarks \cite{ps}.  
  
If Eq.~(\ref{def_width_2.1}) is a consistent definition of width, an important 
question naturally arises: what is the definition of partial widths? 
A recent analysis of that concept, with special emphasis on issues of gauge 
independence and additivity, was given in Ref.~\cite{Grassi:2001dz}.  

The aim of the present paper is to revisit the important problem of width and
partial widths of unstable particles in the light of the Nielsen identities
(NI) \cite{kon,nie,klu,pig,kum}.
Since the variation of Green functions with respect to gauge parameters can be
viewed as Becchi-Rouet-Stora-Tyutin (BRST) transformations, it is
convenient to enlarge the BRST symmetry to include also the gauge parameters
themselves.
In this new framework, known as extended BRST symmetry, a particular set of
Slavnov-Taylor identities \cite{st}, which are frequently called NI, describe
the gauge dependence of Green functions \cite{kon,nie,klu,pig,kum}.
(Originally, the NI were formulated for one-particle-irreducible (1PI) Green
functions in connection with the effective scalar potential in a special class
of gauges.)
In Sec.~II, we employ the NI to show that $M$ and $\Gamma$,  
defined by Eqs.~(\ref{def_mass_1}) and (\ref{def_width_1}), are indeed gauge 
dependent in $O(g^4)$ and $O(g^6)$, respectively. 
For completeness, we also briefly review the proof that $\bar s$ is gauge 
independent \cite{gg}. 
In Sec.~III, we apply the NI to understand and clarify several of 
the results and theoretical issues presented in Ref.~\cite{Grassi:2001dz}. 
In particular, we show that a definition of branching ratios and partial 
widths based on the pole residues is gauge independent to all orders, while a 
simpler, previously discussed formulation is gauge independent through NNLO. 
In Sec.~IV, we explain how the simpler definition can be 
modified to extend the gauge independence to all orders. 
In Sec.~V, we discuss the NI for box diagrams and show that the 
physical amplitude is the most general combination of self-energy, vertex, and 
box contributions that is gauge independent for all values of $s$. 
Section~VI deals with the analytical properties of the NI functions,
and the Appendix presents their one-loop expressions in the $Z$-$\gamma$
sector of the Standard Model (SM). 
  
\boldmath 
\section{NI FOR $\Pi(s,\xi_k)$ AND GAUGE DEPENDENCE OF $M$ AND $M\Gamma$}  
\unboldmath 
  
The transverse propagator of a gauge field is of the form  
\begin{equation} 
\label{new_1}  
{\cal D}_{\mu\nu}=-i\frac{Q_{\mu\nu}}{\Pi(s,\xi_k)},   
\end{equation}  
where $Q_{\mu\nu}=g_{\mu\nu}-p_\mu p_\nu/s$, $p_\mu$ is the four-momentum,    
$s=p^2$, and $\xi_k$ is a generic gauge parameter. 
In the absence of mixing, we have    
\begin{equation}  
\label{new_1.1}  
\Pi(s,\xi_k)=s-M_0^2-A(s,\xi_k) 
=s-m_2^2-\re[A(s,\xi_k)-A(\bar s,\xi_k)]-i\im A(s,\xi_k).  
\end{equation} 
In the last expression, $M_0^2$ has been expressed in terms of $m_2^2$ via 
Eq.~(\ref{def_width_2}).   
  
The NI for $\Pi(s,\xi_k)$ reads
\begin{eqnarray}
\label{ni_1}  
{\partial\over\partial\xi_l}\Pi(s,\xi_k)=2\Lambda_l(s,\xi_k)\Pi(s,\xi_k),  
\end{eqnarray}
where $\Lambda_l(s,\xi_k)$ is a complex, amputated, 1PI, two-point Green
function of $O(g^2)$ involving the gauge field, its BRST variation, and the
gauge fermion (see, e.g., Eq.~(42) of Ref.~\cite{gg}, Eq.~(2.15) of
Ref.~\cite{nie}, Eq.~(4.9a) of Ref.~\cite{klu}, and Eq.~(21) of 
Ref.~\cite{kum}).
We recall that the sum of the gauge-fixing and ghost terms in the Lagrangian 
density can be expressed as the BRST variation of the gauge fermion, and that
the latter is coupled to the BRST variation of the gauge parameter $\xi_l$.
It is understood that the vertex corresponding to the BRST variation of 
$\xi_l$ carries zero momentum, so that $\Lambda_l(s,\xi_k)$ depends  
kinematically only on $s$. 
The simple form of Eq.~(\ref{ni_1}) can only be achieved by a proper treatment 
of the Higgs tadpole: either one chooses to renormalize the one-point function 
to cancel the tadpole contribution or to reabsorb it into a redefinition of 
the self-energy $\Pi(s,\xi_k)$ \cite{gg,si_80}.  
  
Eq.~(\ref{ni_1}) permits one to immediately understand the gauge independence 
of $\bar s$ \cite{gg}. 
As $\bar s$ is the zero of $\Pi(s,\xi_k) $, we have   
\begin{equation}  
\label{ni_3}  
\Pi(\bar s,\xi_k)=0.  
\end{equation}  
Differentiating Eq.~(\ref{ni_3}) with respect to $\xi_l$, we obtain   
\begin{equation} 
\label{ni_4}  
{\partial \bar s \over \partial \xi_l}\, 
{\partial \over \partial \bar s}\Pi(\bar s, \xi_k) +   
{\partial \over \partial \xi_l} \Pi(\bar s, \xi_k) = 0 .   
\end{equation}  
However, Eqs.~(\ref{ni_1}) and (\ref{ni_3}) tell us that the second term in 
Eq.~(\ref{ni_4}) vanishes. 
Since   
$\partial\Pi(\bar s,\xi_k)/\partial\bar s=1+O(g^2)$, we find  
\begin{equation} 
\label{ni_5} 
{\partial \bar s \over \partial \xi_l} = 0,  
\end{equation}  
the result of Ref.~\cite{gg}.   
  
Instead, taking the real part of Eq.~(\ref{ni_1}), we have   
\begin{equation} 
\label{ni_6}  
{\partial\over\partial\xi_l}\re\Pi(s,\xi_k) 
=2\re\Lambda_l(s,\xi_k)\re\Pi(s,\xi_k)-2\im\Lambda_l(s,\xi_k)\im\Pi(s,\xi_k). 
\end{equation} 
By definition, the on-shell mass is the zero of $\re\Pi(s,\xi_k)$. 
Thus, 
\begin{equation} 
\label{ni_7} 
\re\Pi(M^2,\xi_k)=0. 
\end{equation}  
Differentiating Eq.~(\ref{ni_7}) with respect to $\xi_l$ and using 
Eq.~(\ref{ni_6}), we have   
\begin{equation}  
\label{ni_8}  
{\partial M^2\over\partial\xi_l}\re\Pi^\prime(M^2,\xi_k) 
-2\im\Lambda_l(M^2,\xi_k)\im\Pi(M^2,\xi_k)=0,   
\end{equation}  
where the prime indicates a derivative with respect to the first argument. 
Eq.~(\ref{ni_8}) leads to 
\begin{equation} 
\label{ni_9}  
{\partial M^2\over\partial\xi_l} 
=2{\im\Lambda_l(M^2,\xi_k)\im\Pi(M^2,\xi_k)\over\re\Pi^\prime(M^2,\xi_k)}.  
\end{equation}  
Since $\re\Pi^\prime(M^2,\xi_k)=O(1)$ and both $\im\Lambda_l(M^2,\xi_k)$ and 
$\im\Pi(M^2,\xi_k)$ are of $O(g^2)$, Eq.~(\ref{ni_9}) tells us that 
${\partial M^2/\partial\xi_l}=O(g^4)$, the conclusion obtained in the past   
\cite{si_1,psz,si-kniehl,gg}.
  
Turning our attention to Eq.~(\ref{def_width_1}), we note that it may be 
written as 
\begin{equation}  
\label{ni_10}  
M\Gamma={\im\Pi(M^2,\xi_k)\over\re\Pi^\prime(M^2,\xi_k)}.  
\end{equation}  
The imaginary part of Eq.~(\ref{ni_1}) tells us that   
\begin{equation} 
\label{ni_11}  
{\partial\over\partial\xi_l}\im\Pi(M^2,\xi_k)
=2\re\Lambda_l(M^2,\xi_k)\im\Pi(M^2,\xi_k),  
\end{equation}  
while, from Eqs.~(\ref{ni_6}) and (\ref{ni_7}), we see that 
\begin{equation}  
\label{ni_12}  
{\partial\over\partial\xi_l}\re\Pi^\prime(M^2,\xi_k) 
=2\re\Lambda_l(M^2,\xi_k)\re\Pi^\prime(M^2,\xi_k) 
-2\left[\im\Lambda_l(M^2,\xi_k)\im\Pi(M^2,\xi_k)\right]^\prime.  
\end{equation} 
Combining Eqs.~(\ref{ni_11}) and (\ref{ni_12}), we have   
\begin{equation}  
\label{ni_13}  
{\partial\over\partial\xi_l}\,{\im\Pi(M^2,\xi_k)\over\re\Pi^\prime(M^2,\xi_k)} 
=2{\im\Pi(M^2,\xi_k)[\im\Lambda_l(M^2,\xi_k)\im\Pi(M^2,\xi_k)]^\prime\over 
[\re\Pi^\prime(M^2,\xi_k)]^2}.  
\end{equation} 
To obtain the total derivative, we add the term   
$({\partial M^2/\partial\xi_l})\partial[{\im\Pi(M^2)/\re\Pi^\prime(M^2)}]
/\partial M^2$ 
and find that   
\begin{equation} 
\label{ni_14}  
{d\over d\xi_l}\,{\im\Pi(M^2)\over\re\Pi^\prime(M^2)} 
=2{\{\im\Lambda_l(M^2)[\im\Pi(M^2)]^2\}^\prime\over[\re\Pi^\prime(M^2)]^2} 
-2{\im\Lambda_l(M^2)[\im\Pi(M^2)]^2\re\Pi^{\prime\prime}(M^2)\over 
[\re\Pi^\prime(M^2)]^3}.   
\end{equation}  
In Eq.~(\ref{ni_14}) and henceforth, we do not explicitly display the
dependence of the various Green functions on $\xi_k$. 

Since $\im\Lambda_l(M^2)$, $\im\Pi(M^2)$, and $\re\Pi^{\prime\prime}(M^2)$ are 
of $O(g^2)$ and $\re\Pi^\prime(M^2)=O(1)$, in leading order, Eq.~(\ref{ni_14}) 
reduces to 
\begin{equation}  
\label{ni_15}  
{d\over d\xi_l}\,{\im\Pi(M^2)\over\re\Pi^\prime(M^2)} 
=2\left\{\im\Lambda_l(M^2)[\im\Pi(M^2)]^2\right\}^\prime+O(g^8).  
\end{equation}  
Thus, we see that the conventional definition of width 
[Eq.~(\ref{def_width_1})] is gauge dependent in NNLO, i.e.\ in $O(g^6)$, in 
agreement with the conclusion of Refs.~\cite{si_1,psz,si-kniehl,gg}.  
  
An important implication of the above results is that, in a gauge theory, 
Eqs.~(\ref{def_mass_1}) and (\ref{def_width_1}) can be identified with the 
physical observables only through next-to-leading order (NLO).  
  
In the case of mixing between two fields $A$ and $B$, Eq.~(\ref{ni_1}) is   
replaced by  
\begin{eqnarray}\label{ni_2}  
{\partial\over\partial\xi_l}\Pi_{\alpha\beta}(s) 
=\sum_\delta\left[\Lambda_{l,\alpha}^\delta(s)\Pi_{\delta\beta}(s) 
+\Lambda_{l,\beta}^\delta(s)\Pi_{\delta\alpha}(s)\right],   
\end{eqnarray}  
where $\alpha,\beta,\delta=A,B$, the diagonal elements 
$\Pi_{\alpha\alpha}(s)\equiv s-M_{0,\alpha}^2-A_{\alpha\alpha}(s)$ are 
the analogues of Eq.~(\ref{new_1.1}), and, for $\alpha\neq\beta$,   
$\Pi_{\alpha\beta}(s)\equiv-A_{\alpha\beta}(s)$, with $A_{\alpha\beta}(s)$ 
being the mixed $A$-$B$ self-energy (we use the sign conventions of 
Ref.~\cite{si_80}). 
Here, $\Lambda_{l,\alpha}^\delta(s)$ are mixed two-point functions involving 
the field $\alpha$, the BRST variation of $\xi_l$, and the source of the BRST 
variation of the field $\delta$. 
Correspondingly, the pole positions $\bar s_A$ and $\bar s_B$ are the zeroes 
of   
\begin{equation}  
\label{pole_mixed}  
D(s)=\Pi_{AA}(s)\Pi_{BB}(s)-\Pi^2_{AB}(s).  
\end{equation}  
  
Using Eq.~(\ref{ni_2}), one finds  
\begin{equation} 
\label{pole_mixed_1}  
{\partial\over\partial\xi_l}D(s) 
=2\left[\Lambda_{l,A}^A(s)+\Lambda_{l,B}^B(s)\right]D(s),  
\end{equation} 
and, therefore, 
\begin{equation}  
\label{pole_mixed_2}  
{\partial\over\partial\xi_l}D(\bar s)=0.   
\end{equation}  
  
Differentiating $D(\bar s)=0$ with respect to $\xi_l$ and employing 
Eq.~(\ref{pole_mixed_2}), one obtains again the result 
${\partial\bar s/\partial\xi_l}=0$ \cite{gg}.   
  
The transverse propagator of the field $A$ is   
\begin{equation}  
\label{prop_1} 
-i{\Pi_{BB}(s)\over D(s)}={-i\over s-M_{0,A}^2-A(s)},   
\end{equation}  
where 
\begin{equation}  
\label{prop_2}  
A(s)=A_{AA}(s)+{A_{AB}^2(s)\over\Pi_{BB}(s)}.   
\end{equation}  
We also note that Eqs.~(\ref{ni_2}) and (\ref{pole_mixed_1}) lead to   
\begin{eqnarray}  
\label{prop_3} 
{\partial\over\partial\xi_l}\,{D(s)\over\Pi_{BB}(s)} 
=2\left[\Lambda_{l,A}^A(s)-{\Lambda_{l,B}^A(s)\Pi_{AB}(s)\over\Pi_{BB}(s)} 
\right]{D(s)\over\Pi_{BB}(s)}.    
\end{eqnarray}  
  
\section{NI FOR VERTEX FUNCTIONS AND GAUGE PROPERTIES OF POLE RESIDUES AND
PARTIAL WIDTHS}
  
We consider the amplitude $i\to Z\to f$, where $Z$ is an unstable gauge boson, 
and $i$ and $f$ are initial and final states, respectively, involving on-shell
particles that are either stable or have negligible widths.   
  
Using Eq.~(\ref{def_mass_2}), we may express Eq.~(\ref{new_1}) as   
\begin{equation}  
\label{res_1}  
{\cal D}_{\mu\nu}=-i\frac{Q_{\mu\nu}}{s-\bar s-A(s)+A(\bar s)}.   
\end{equation}  
The vertex amplitude defined by $Z$ and $f$ is given by 
\begin{equation}  
\label{res_2}  
V_f^\mu(s)\equiv\left\langle f\left|J_Z^\mu\right|0\right\rangle  
=\sum_av_f^{(a)}(s,\dots)M_f^{(a)\mu},  
\end{equation}  
where $M_f^{(a)\mu}$ denote various independent vector and axial-vector matrix  
elements involving the spinors, polarization four-vectors, and four-momenta of  
the final-state particles, while $v_f^{(a)}(s,\dots)$ are scalar functions.  
Here, it is understood that the amplitude $V_f^\mu(s)$ is not 1PI, but
includes the field renormalizations of the external, on-shell particles in the
state $f$.
The dots indicate their dependence on the additional invariants constructed 
from the particles' momenta.   
We use the convention of including the generic coupling $g$ in the definition 
of $v_f^{(a)}(s,\dots)$, so that, in leading order, $v_f^{(a)}(s,\dots)=O(g)$.
Expanding $V_f^\mu(s)$ and the denominator of Eq.~(\ref{res_1}) about 
$s=\bar s$, the overall amplitude may be expressed as   
\begin{equation}\label{res_3}  
{\cal A}_{fi}(s)=-i\frac{Q_{\mu\nu}V_f^\mu(\bar s)V_i^\nu(\bar s)}  
{(s-\bar s)[1-A^\prime(\bar s)]}+N_{fi},  
\end{equation}  
where $N_{fi}$ stands for non-resonant contributions and, henceforth, we do 
not indicate the dependence on the additional invariants.   
  
In the absence of mixing and using the extended BRST formalism, it is possible
to derive the following identity for the gauge dependence of $V_f^\mu(s)$:
\begin{equation}
\label{res_4}
{\partial\over\partial\xi_l}V_f^\mu(s) 
=\Lambda_l(s)V_f^\mu(s)+\Pi(s)\Delta_{l,f}^\mu(s),  
\end{equation}  
where $\Delta_{l,f}^\mu(s)$ is a complex Green function depending on the
scalar invariants and involving the gauge field $Z$, the gauge fermion, and
the sources for the fields of the final state $f$ (see, e.g., Eq.~(45) of
Ref.~\cite{gg}, Eq.~(4.9b) of Ref.~\cite{klu}, and Eq.~(14) of
Ref.~\cite{kum}).
For instance, in a four-fermion amplitude, $f$ coincides with the two outgoing
on-shell fermions, and, therefore, $\Delta_{l,f}^\mu(s)$ involves the latter. 
Although, as discussed after Eq.~(\ref{ni_1}), the function
$\Lambda_l(s,\xi_k)$ is a 1PI Green function, the new Green function
$\Delta_{l,f}^\mu(s)$ is of the same nature as $V_f^\mu(s)$.
It is functionally independent of $\Lambda_l(s)$ and incorporates the gauge 
dependence of the vertex $V_f^\mu(s)$ that is not proportional to the vertex 
itself. 
In leading order, $\Delta_{l,f}^\mu(s)=O(g^3)$.
It is important to emphasize that the function $\Delta_{l,f}^\mu(s)$ in
Eq.~(\ref{res_4}) does not contain contributions proportional to $1/\Pi(s)$
and, in fact, is not singular at $s=\bar s$.

Differentiating Eq.~(\ref{ni_1}) with respect to $s$ and setting $s=\bar s$, 
we have  
\begin{equation} 
\label{res_6} 
{\partial\over\partial\xi_l}\Pi^\prime(\bar s) 
=2\Lambda_l(\bar s)\Pi^\prime(\bar s),  
\end{equation}  
while Eq.~(\ref{res_4}) leads to 
\begin{equation} 
\label{res_7} 
{\partial\over\partial\xi_l}V_f^\mu(\bar s) 
=\Lambda_l(\bar s)V_f^\mu(\bar s).   
\end{equation}  
Combining Eqs.~(\ref{res_6}) and (\ref{res_7}), and recalling that $\bar s$ is 
$\xi_l$ independent, we obtain 
\begin{equation} 
\label{res_8} 
{d\over d\xi_l}{V_f^\mu(\bar s)\,\over\sqrt{1-A^\prime(\bar s)}}=0,  
\end{equation}  
for any choice of $f$, a result that implies the gauge independence of the 
pole residues in Eq.~(\ref{res_3}). 
This is expected on general grounds \cite{si_1,zmass}, since such amplitudes 
are the residues of poles in the analytically continued scattering ($S$)
matrix. 
On the other hand, Eq.~(\ref{res_8}) provides a formal proof of this important 
conclusion.   
  
In the case of mixing between two fields $A$ and $B$, Eq.~(\ref{res_4}) is 
generalized to   
\begin{equation}  
\label{res_5} 
{\partial\over\partial\xi_l}\Gamma_{\alpha,f}^\mu(s) 
=\sum_\delta\left[\Lambda_{l,\alpha}^\delta(s)\Gamma_{\delta,f}^\mu(s) 
+\Pi_{\alpha\delta}(s)\Delta_{l,f}^{\delta,\mu}(s)\right],   
\end{equation}  
where $\alpha,\delta=A,B$ and $\Gamma_{\delta,f}^\mu(s)$ stand for vertex
amplitudes that, again, include the field renormalizations of the external, 
on-shell particles. 
However, in the mixing case, the relevant vertex parts are combinations of the 
form  
\begin{equation}  
\label{new_ver_1}  
V^\mu_{A,f}(s)=\Gamma_{A,f}^\mu(s) 
-{\Pi_{AB}(s)\Gamma^\mu_{B,f}(s)\over\Pi_{BB}(s)}.  
\end{equation}  
For instance, in the neutral-current amplitude of the SM, the second term in 
Eq.~(\ref{new_ver_1}) corresponds to one-particle-reducible contributions to 
the $Z^0f\bar f$ vertex arising from the $Z$-$\gamma$ mixing.   
Eqs.~(\ref{ni_2}), (\ref{res_5}), and (\ref{new_ver_1}) lead to   
\begin{equation}  
\label{new_ver_2}  
{\partial\over\partial\xi_l}V_{A,f}^\mu(s) 
=\left[\Lambda_{l,A}^A(s) 
-{\Lambda_{l,B}^A(s)\Pi_{AB}(s)\over\Pi_{BB}(s)}\right]V_{A,f}^\mu(s) 
+{D(s)\over\Pi_{BB}(s)}\left[\Delta_{l,f}^{A,\mu}(s) 
-{\Lambda_{l,B}^A(s)\Gamma_{B,f}^\mu(s)\over\Pi_{BB}(s)}\right].   
\end{equation}  
Setting $s=\bar s$, Eq.~(\ref{new_ver_2}) reduces to   
\begin{equation}  
\label{new_ver_3}  
{\partial\over\partial\xi_l}V_{A,f}^\mu(\bar s) 
=\left[\Lambda_{l,A}^A(\bar s)
-{\Lambda_{l,B}^A(\bar s)\Pi_{AB}(\bar s)\over\Pi_{BB}(\bar s)}\right] 
V_{A,f}^\mu(\bar s),  
\end{equation} 
which is the generalization of Eq.~(\ref{res_7}). 
Differentiating Eq.~(\ref{prop_3}) with respect to $s$, setting $s =\bar s$, 
and using Eq.~(\ref{new_ver_3}), we find   
\begin{equation} 
\label{new_ver_4}  
{d\over d\xi_l}\,{V_{A,f}^\mu(\bar s) 
\over\sqrt{[D(s)/\Pi_{BB}(s)]_{s=\bar s}^\prime}}=0,  
\end{equation}  
for any choice of $f$. 
Noting that $[D(s)/\Pi_{BB}(s)]_{s=\bar s}^\prime=1-A^\prime(\bar s)$  
[cf.\ Eqs.~(\ref{prop_1}) and (\ref{prop_2})], we see that 
Eq.~(\ref{new_ver_4}) generalizes Eq.~(\ref{res_8}) and implies the gauge   
independence of the pole residues in the mixed case.    
  
The second term on the r.h.s.\ of Eq.~(\ref{new_ver_2}) informs us that, in 
the case of mixing, loop corrections generate $O(g^3)$ gauge-dependent 
contributions to $V_{A,f}^\mu(s)$ that are not proportional to the 
lowest-order matrix elements $V_{A,f}^{\mu(0)}(s)$.
(Here and in the following, a superfix $(n)$ indicates that the respective
quantity is considered at $n$ loops.)
They are proportional to the inverse propagator 
$D(s)/\Pi_{BB}(s)=\Pi_{AA}(s)-\Pi_{AB}^2(s)/\Pi_{BB}(s)$ and, therefore, 
vanish at $s=\bar s$. 
This is necessary, since they cannot be cancelled by other $O(g^3)$ 
contributions in the amplitude 
$V_{A,f}^\mu(\bar s)/\sqrt{[D(s)/\Pi_{BB}(s)]_{s=\bar s}^\prime}$. 
On the other hand, the proportionality to the inverse propagator implies that 
they contribute to the non-resonant amplitude, where they cancel other 
gauge-dependent terms. 
Well-known examples are contributions of this type to the neutral-current 
amplitudes of the SM involving the $Z$-$\gamma$ self-energy $A_{Z\gamma}(s)$, 
which cancel gauge-dependent terms in the photon-mediated amplitude and box 
diagrams \cite{deg}. 
We will refer to these terms as off-diagonal contributions.   
  
The above-mentioned property of the residues has motivated a gauge-independent   
definition of partial width, to wit \cite{Grassi:2001dz}  
\begin{equation} 
\label{res_9} 
m_2\hat\Gamma_f=-{1\over 6}\sum_{\rm spins}\int d\Phi_f   
\frac{Q_{\mu\nu}V_f^{\mu*}(\bar s)V_f^\nu(\bar s)}  
{\left|1-A^\prime(\bar s)\right|},  
\end{equation}  
where the integration is over the phase space of the final-state particles, a 
factor of $1/3$ arises from the average over the initial-state polarization, 
and a factor of $1/2$ from the familiar relation between $m_2\hat\Gamma_f$ and 
the integrated, squared amplitude.   
In the mixing case, it is understood that $V_f^\mu(s)$ is defined according to   
Eq.~(\ref{new_ver_1}) and $A(s)$ according to Eq.~(\ref{prop_2}).   
  
In Ref.~\cite{Grassi:2001dz}, it was pointed out that  
$\sum_f\hat\Gamma_f\neq\Gamma_2$ when NNLO contributions are included, i.e.\ 
the sum of partial widths defined by means of Eq.~(\ref{res_9}) does not add 
up to the total width $\Gamma_2$ defined on the basis of the pole position 
$\bar s$ [Eq.~(\ref{def_width_2.1})]. 
In principle, the lack of additivity can be circumvented by a simple 
rescaling: one defines the branching ratios by 
$B_f=\hat\Gamma_f/\sum_f\hat\Gamma_f$ and redefines the partial width as 
$\Gamma_f=B_f\Gamma_2$, so that $\sum_f\Gamma_f=\Gamma_2$
\cite{Grassi:2001dz}.  
  
An analysis that leads to an alternative definition of partial widths, much 
closer to the conventional one, is based on the overall amplitude evaluated at 
$s=m_2^2$ \cite{Grassi:2001dz}, 
\begin{equation}\label{res_10} 
{\cal A}_{fi}\left(m_2^2\right) 
=-i\frac{Q_{\mu\nu}V_f^\mu\left(m_2^2\right)V_i^\nu(m_2^2)}  
{im_2\Gamma_2-A\left(m_2^2\right)+A\left(\bar s\right)}+\tilde{N}_{fi},  
\end{equation}  
where again $\tilde{N}_{fi}$ represents non-resonant contributions. 
It is understood that the gauge-dependent, off-diagonal terms proportional to  
the inverse propagator 
$D\left(m_2^2\right)/\Pi_{BB}\left(m_2^2\right)\approx im_2\Gamma_2$, 
discussed in the paragraph after Eq.~(\ref{new_ver_4}), are not included in 
$V_f^{\mu}\left(m_2^2\right)$ or $V_i^{\nu}\left(m_2^2\right)$, since they are 
needed to cancel other contributions of the same type in the non-resonant part 
$\tilde{N}_{fi}$.  
  
Eq.~(\ref{res_10}) suggests the consideration of the amplitudes   
\begin{equation}  
\label{res_11}  
I_f\left(m^2_2\right)=\frac{1}{6}\sum_{\rm spins}\int d\Phi_f   
Q_{\mu\nu}V_f^{\mu*}\left(m_2^2\right)V_f^\nu\left(m_2^2\right).  
\end{equation}  
In fact, $-I_f\left(m^2_2\right)/m_2$ is the conventional expression for the 
partial width of the decay of the unstable particle into the physical state  
$f$, modulo its wave-function renormalization. 
In particular, $I_f\left(m^2_2\right)$ is the $f$-cut contribution to 
$I\left(m_2^2\right)\equiv\im A\left(m_2^2\right)$ in the decomposition 
explained after Eq.~(\ref{def_width_1}), except that it is evaluated at the 
gauge-independent pole mass $m_2$, rather than the on-shell mass $M$. 
As explained in Ref.~\cite{Grassi:2001dz}, if the unstable particle were an 
asymptotic state, the unitarity of the $S$ matrix would imply that 
$I\left(m_2^2\right)=F\left(m_2^2\right)$, where  
\begin{equation}  
\label{res_12}  
F\left(m_2^2\right)\equiv\sum_fI_f\left(m^2_2\right).  
\end{equation}  
  
Since the unstable particle is not an asymptotic state, this is not the case 
beyond NLO, and we expect a relation of the form   
\begin{equation}  
\label{res_13}  
I\left(m_2^2\right)=F\left(m_2^2\right)+G\left(m_2^2\right),  
\end{equation}  
where $G\left(m_2^2\right)\equiv I\left(m_2^2\right)-F\left(m_2^2\right)$ 
involves contributions from unphysical intermediate states, including would-be 
Goldstone bosons, Faddeev-Popov ghosts, or longitudinal modes of gauge bosons. 
By studying the gauge-independent difference 
$m_2\sum_f\hat\Gamma_f-m_2\Gamma_2$, it was shown in Ref.~\cite{Grassi:2001dz} 
that $G\left(m_2^2\right)$ is non-vanishing in $O(g^6)$, i.e.\ in NNLO.  
 
A definition of branching ratios, very similar to the conventional one, was 
discussed in Ref.~\cite{Grassi:2001dz}, to wit 
\begin{equation}  
\label{res_14} 
\tilde B_f=\frac{I_f\left(m_2^2\right)}{F\left(m_2^2\right)}.   
\end{equation}  
It is manifestly additive: $\sum_f\tilde B_f=1$. 
The partial widths are then defined by   
\begin{equation}    
\label{res_15}  
\tilde\Gamma_{f}=\tilde B_f\Gamma_2.  
\end{equation}  
In order to examine the gauge dependence of $I_f\left(m_2^2\right)$, we employ 
the NI of Eq.~(\ref{res_4}), with $s=m_2^2$, and find   
\begin{eqnarray} 
\label{res_15.1} 
{\partial\over\partial\xi_l}I_f\left(m_2^2\right) 
&=&2\re\Lambda_l\left(m_2^2\right)I_f\left(m_2^2\right) 
+{1\over6}\sum_{\rm spins}\int d\Phi_fQ_{\mu\nu}   
\left[V_f^{\mu*}\left(m_2^2\right)\Pi\left(m_2^2\right) 
\Delta_{l,f}^\nu\left(m_2^2\right)\right. 
\nonumber\\ 
&&{}+\left.\Delta_{l,f}^{\mu*}\left(m_2^2\right)\Pi^*\left(m_2^2\right) 
V^\nu_f\left(m_2^2\right)\right].  
\end{eqnarray}  
In leading order, we have $\Pi(m_2^2)=-iI\left(m_2^2\right)$ and   
$\Delta_{l,f}^\nu\left(m_2^2\right) 
=\delta_l\left(m_2^2\right)V_f^\nu\left(m_2^2\right)$, where   
$\delta_l\left(m_2^2\right)=O(g^2)$ is independent of the final state $f$. 
An illustration of this property is provided by Eq.~(\ref{res_23}) and by the 
coefficient of $\Gamma_{Z,f}^{\mu(0)}(s)$ on the r.h.s.\ of Eq.~(\ref{app_6}). 
Thus, in leading order, Eq.~(\ref{res_15.1}) becomes   
\begin{equation} 
\label{res_16} 
{\partial\over\partial\xi_l}I_f\left(m_2^2\right) 
=2\left[\re\Lambda_l\left(m_2^2\right)+ 
I\left(m_2^2\right)\im\delta_l\left(m_2^2\right)\right]I_f\left(m_2^2\right) 
+O(g^8).  
\end{equation}  
  
We note that $\Lambda_l\left(m_2^2\right)$, $I\left(m_2^2\right)$, 
$\delta_l\left(m_2^2\right)$, and $I_f\left(m_2^2\right)$ are of $O(g^2)$. 
Eq.~(\ref{res_16}) tells us that: (i) the gauge dependence of  
$I_f\left(m_2^2\right)$ starts in $O(g^4)$ (NLO), a well-known fact; and 
(ii) through $O(g^6)$ (NNLO), $\partial I_f\left(m_2^2\right)/\partial\xi_l$ 
is proportional to $I_f\left(m_2^2\right)$, with an $f$-independent 
coefficient. 
In turn, this proportionality implies the gauge independence of $\tilde B_f$ 
[Eq.~(\ref{res_14})] through NNLO! 
The same conclusion holds in the presence of mixing, provided that, as 
explained after Eq.~(\ref{res_10}), gauge-dependent, off-diagonal terms 
proportional to $im_2\Gamma_2$ are not included in 
$V_f^\mu\left(m_2^2\right)$.   
  
This observation explains a significant result of Ref.~\cite{Grassi:2001dz}, 
namely it was shown in that work that, when the cross section of $e^+e^-$ 
annihilation at the $Z^0$-boson peak in the SM is expressed in terms of the 
partial widths $\tilde\Gamma_f$, it is gauge independent through NNLO. 
From this conclusion it was inferred in Ref.~\cite{Grassi:2001dz} that the 
partial widths $\tilde\Gamma_f$ are also gauge independent through NNLO, i.e.\ 
through $O(g^6)$. 
We now see that this result is related to the proportionality between  
$\partial I_f\left(m_2^2\right)/\partial\xi_l$ and $I_f\left(m_2^2\right)$ 
with an $f$-independent coefficient, exhibited in Eq.~(\ref{res_16}) through  
$O(g^6)$. 
On the other hand, there is no reason to expect that such a proportionality 
survives in still higher orders.  
  
Thus, although the definitions of Eqs.~(\ref{res_14}) and (\ref{res_15}) 
provide gauge-independent results through NNLO, and this is sufficient for the 
phenomenological requirements of electroweak physics in the foreseeable 
future, they are not completely satisfactory from the theoretical point of 
view, since they are not expected to be gauge independent in still higher 
orders.  
  
Next, we apply the NI to obtain information about the function  
$G\left(m_2^2\right)$ introduced in Eq.~(\ref{res_13}). 
Taking the imaginary part of Eq.~(\ref{ni_1}), setting $s=m_2^2$, and 
recalling from Eq.~(\ref{new_1.1}) that   
$\im\Pi\left(m_2^2\right)=-I\left(m_2^2\right)$, we have   
\begin{equation} 
\label{res_17} 
{\partial\over\partial\xi_l}I\left(m_2^2\right) 
=2\re\Lambda_l\left(m_2^2\right)I\left(m_2^2\right) 
+2\im\Lambda_l\left(m_2^2\right)\re\left[A\left(m_2^2\right)-A(\bar s)\right].
\end{equation} 
Through $O(g^6)$, this becomes   
\begin{equation} 
\label{res_18} 
{\partial\over\partial\xi_l}I\left(m_2^2\right) 
=2\re\Lambda_l\left(m_2^2\right)I\left(m_2^2\right) 
-2m_2\Gamma_2\im\Lambda_l\left(m_2^2\right)I^\prime\left(m_2^2\right)+O(g^8). 
\end{equation} 
Inserting Eq.~(\ref{res_13}) into Eq.~(\ref{res_18}) and subtracting   
\begin{equation} 
\label{res_19} 
{\partial\over\partial\xi_l}F\left(m_2^2\right) 
=2\left[\re\Lambda_l\left(m_2^2\right) 
+I\left(m_2^2\right)\im\delta_l\left(m_2^2\right)\right]F\left(m_2^2\right), 
\end{equation}  
a relation that follows from Eq.~(\ref{res_16}), we find 
\begin{equation} 
\label{res_20} 
{\partial\over\partial\xi_l}G\left(m_2^2\right) 
={m_2\Gamma_2\over2}\,{\partial\over\partial\xi_l}   
\left[I^\prime\left(m_2^2\right)\right]^2 
-2m_2^2\Gamma_2^2\im\delta_l\left(m_2^2\right)+O(g^8). 
\end{equation}  
In Eq.~(\ref{res_20}), we have neglected terms of $O(g^8)$ and employed   
\begin{equation} 
\label{res_21} 
\im\Lambda_l\left(m_2^2\right) 
=-\frac{1}{2}{\partial\over\partial\xi_l}I^\prime\left(m_2^2\right)+O(g^4), 
\end{equation}  
a result that follows from Eq.~(\ref{ni_1}). 
Eq.~(\ref{res_20}) tells us that $G\left(m_2^2\right)$ differs from zero in 
$O(g^6)$ (NNLO), a conclusion also reached by a different argument in  
Ref.~\cite{Grassi:2001dz}. 
In fact, Eq.~(16) of Ref.~\cite{Grassi:2001dz} presents an explicit expression 
for the $O(g^6)$ contribution to $G\left(m_2^2\right)$ in the $Z^0$-boson 
case of the SM, obtained by considering the difference between the two 
gauge-independent amplitudes $m_2\sum_f\hat \Gamma_f$ and $m_2\Gamma_2$. 
As shown in that expression, in the $Z^0$-boson case, the leading contribution 
to $G\left(m_2^2\right)$ depends on the gauge parameter $\xi_W$ associated  
with the $W$ boson. 
Differentiating Eq.~(16) of Ref.~\cite{Grassi:2001dz} with respect to $\xi_W$, 
we find 
\begin{equation} 
\label{res_22} 
{\partial\over\partial\xi_W}G\left(m_2^2\right) 
={m_2\Gamma_2\over 2}\,{\partial\over\partial\xi_W} 
\left[I^\prime\left(m_2^2\right)\right]^2   
-g^2c_w^2m_2^2\Gamma_2^2{\partial\over\partial\xi_W}   
\left[(\xi_W-1)\im\eta_W\left(m_2^2\right)\right]+O(g^8),  
\end{equation}  
where $c_w\equiv\cos\theta_w$, with $\theta_w$ being the weak mixing angle,
and $\eta_W(s)$ is a gauge-dependent amplitude introduced in Ref.~\cite{deg}.
Comparison of Eqs.~(\ref{res_20}) and (\ref{res_22}) shows that the two 
expressions have the same structure with the identification   
\begin{equation} 
\label{res_23} 
\im\delta_W\left(m_2^2\right)=\frac{g^2c_w^2}{2}\,{\partial\over\partial\xi_W}  
\left[(\xi_W-1)\im\eta_W\left(m_2^2\right)\right]. 
\end{equation}  
Thus, the NI permit us to understand the non-vanishing of 
$G\left(m_2^2\right)$ and the structure of its leading contribution. 
As shown in Ref.~\cite{Grassi:2001dz}, the contribution to 
$G\left(m_2^2\right)$ involving $\im\eta_W\left(m_2^2\right)$ plays a crucial 
role in ensuring the gauge independence of the peak cross section through 
NNLO. 
In fact, it cancels the gauge dependence of the interference between the 
resonant amplitude and the box diagrams!  
   
\section{EXTENSION TO ALL ORDERS OF THE GAUGE INDEPENDENCE OF
EQS.~(3.18) AND (3.19)}
  
A strategy to extend to all orders the gauge independence of 
Eq.~(\ref{res_14}), and therefore Eq.~(\ref{res_15}), involves a redefinition 
of the vertex parts $V_f^\mu\left(m_2^2\right)$.
In general, $V_f^\mu(s)$ may be regarded as a function   
$V_f^\mu\left(s,m_2^2,m_2\Gamma_2\right)$. 
The explicit dependence on $m_2\Gamma_2$ arises from specific cut
contributions, or, equivalently, from terms involving $\im A(s)$.
As an illustration, we may consider two-loop contributions to the $Z$-$\gamma$
self-energy $A_{Z\gamma}(s)$ involving virtual fermion-antifermion pairs with
a vertex correction attached to the photon ending.
A cut across the fermion-antifermion pairs, summed over all pairs, leads at 
$s=m_2^2$ to a contribution proportional to $im_2\Gamma_2$. 
The fact that higher-order corrections generate contributions involving 
$m_2\Gamma_2$ may also be inferred more generally by examining the structure 
of Eqs.~(\ref{prop_3}) and (\ref{new_ver_2}). 
We note that the terms involving $D(s)/\Pi_{BB}(s)$ in these equations are
proportional to $s-\bar s=s-m_2^2+im_2\Gamma_2$, and it is clear that
$im_2\Gamma_2$ must be induced by contributions of higher order than those
leading to $s-m_2^2$. 
In the notation of this section, the amplitudes $V_f^\mu(\bar s)$ and 
$V_f^\mu\left(m_2^2\right)$ are identified with   
$V_f^\mu(\bar s)=V_f^\mu\left(\bar s,m_2^2,m_2\Gamma_2\right)$ and 
$V_f^\mu\left(m_2^2\right)=V_f^\mu\left(m_2^2,m_2^2,m_2\Gamma_2\right)$,   
respectively. 
A modification that renders Eqs.~(\ref{res_14}) and (\ref{res_15}) gauge 
independent to all orders involves replacing $V_f^\mu\left(m_2^2\right)$ in 
Eq.~(\ref{res_11}) by 
$\hat V_f^\mu\left(m_2^2\right)\equiv\lim_{\Gamma_2\to0} 
V_f^\mu\left(\bar s,m_2^2,m_2\Gamma_2\right) 
=V_f^\mu\left(m_2^2,m_2^2,0\right)$. 
We emphasize that $\hat V_f^\mu\left(m_2^2\right)$ differs from the usual 
definition of $V^\mu_f(m_2^2)$. 
Setting $s=m_2^2$ in Eq.~(\ref{new_ver_2}) and taking the limit  
$\Gamma_2\to0$, we see that   
\begin{equation} 
\label{ext_1} 
{\partial\over\partial\xi_l}\hat V_{A,f}^\mu\left(m_2^2\right) 
=\left[\Lambda_{l,A}^A\left(m_2^2\right)
-{\Lambda_{l,B}^A\left(m_2^2\right)\Pi_{AB}\left(m_2^2\right) 
\over\Pi_{BB}\left(m_2^2\right)}\right]_{\Gamma_2=0}   
\hat V_{A,f}^\mu\left(m_2^2\right).   
\end{equation} 
The crucial point is that, in the limit $\Gamma_2\to0$, the second term on the 
r.h.s.\ of Eq.~(\ref{new_ver_2}) vanishes. 
This has the effect of excluding from $\hat V_{A,f}^\mu\left(m_2^2\right)$ all 
the gauge-dependent (diagonal as well as off-diagonal) contributions that are 
proportional to $m_2\Gamma_2$. 
The physical meaning of this exclusion is discussed later.   
 
Replacing $I_f\left(m_2^2\right)$ in Eq.~(\ref{res_11}) by   
\begin{equation} 
\label{ext_2} 
\hat I_f\left(m^2_2\right)=\frac{1}{6}\sum_{\rm spins}\int d\Phi_f   
Q_{\mu\nu}\hat V_f^{\mu*}\left(m_2^2\right)\hat V_f^\nu\left(m_2^2\right),  
\end{equation}  
Eq.~(\ref{ext_1}) leads to the proportionality between 
$\partial\hat I_f\left(m^2_2\right)/\partial\xi_l$ and 
$\hat I_f\left(m^2_2\right)$ with an $f$-indepen\-dent coefficient. 
In turn, this implies the gauge independence to all orders of the modified 
definitions   
\begin{eqnarray} 
\label{ext_3} 
\hat B_f&=&{\hat I_f\left(m_2^2\right)\over\hat F\left(m_2^2\right)}, 
\nonumber\\  
\hat\Gamma_f&=&\hat B_f\Gamma_2,   
\end{eqnarray}  
where $\hat F\left(m_2^2\right)\equiv\sum_f\hat I_f\left(m_2^2\right)$. 
 
A corresponding gauge-independent definition of residues is obtained by 
considering 
\begin{equation} 
\label{ext_4} 
\lim_{\Gamma_2\to0}    
{Q_{\mu\nu}V_f^\mu(\bar s)V_f^\nu(\bar s)\over1-A^\prime(\bar s)} 
={Q_{\mu\nu}\hat V_f^\mu\left(m_2^2\right)\hat V_f^\nu\left(m_2^2\right)\over 
[1-A^\prime(\bar s)]_{\Gamma_2=0}}.   
\end{equation}  
The gauge independence of Eq.~(\ref{ext_4}) follows by taking the limit 
$\Gamma_2\to0$ of Eq.~(\ref{res_8}). 
  
In terms of Eq.~(\ref{ext_4}), the overall  
amplitude of Eq.~(\ref{res_3}) may be expressed as   
\begin{equation} 
\label{ext_5}  
{\cal A}_{fi}(s)=-i\frac{Q_{\mu\nu}\hat V_f^\mu\left(m_2^2\right) 
\hat V_i^\nu\left(m_2^2\right)}{(s-\bar s)[1-A^\prime(\bar s)]_{\Gamma_2=0}} 
+\hat N_{fi}.  
\end{equation}  
  
The physical meaning of the limit $\Gamma_2\to0$ can be easily understood by 
comparing the first terms in Eqs.~(\ref{res_3}) and (\ref{ext_5}). 
The residues of $1/(s-\bar s)$ in the two expressions differ by terms of 
$O(g^4\Gamma_2/m_2)$ (NNLO). 
Such terms give contributions of zeroth order in $\Gamma_2$ to the peak 
amplitude ($s=m_2^2$) and, therefore, may be regarded as non-resonant. 
It should be stressed that, in this approach, terms of 
$O\left((\Gamma_2/m_2)^n\right)$, where $n\ge1$, are not neglected, but they 
are rather incorporated in the non-resonant amplitude.    
Thus, the two expressions differ, in a gauge-independent manner, in the 
precise identification of resonant and non-resonant contributions. 
Although the formulation of Eq.~(\ref{res_3}) is probably more elegant, that 
of Eq.~(\ref{ext_5}) is closer to the calculations carried out by most 
particle physicists.   
  
\section{NI FOR BOX FUNCTIONS}
  
The gauge independence of the complete amplitude can be tested by considering 
the NI for the box functions $B_{fi}(s)$.   
In the absence of mixing, we have    
\begin{equation} 
\label{box_1} 
{\partial\over\partial\xi_l}B_{fi}(s) 
=iQ_{\mu\nu}\left[\Delta_{l,f}^\mu(s)V_i^\nu(s) 
+V_f^\mu(s)\Delta_{l,i}^\nu(s)\right] 
\end{equation}  
(see, e.g., Eq.~(46) of Ref.~\cite{gg} and Eq.~(14) of Ref.~\cite{kum}).
Eq.~(\ref{box_1}) only involves the functions $\Delta_{l,f}^\mu(s)$ appearing
in the NI for vertex functions [Eq.~(\ref{res_4})]. 
This is due to BRST symmetry and the fact that the external states $i$ and $f$
are on shell. 
Indeed, in the NI for off-shell Green functions, there are additional 
contributions proportional to the field equations .   
 
Knowing the gauge dependence of all essential building blocks, we can find the 
most general combination of self-energy, vertex, and box contributions that is
gauge independent for any value of $s$.   
From the NI for self-energies [Eq.~(\ref{ni_1})], we can compute the function 
$\Lambda_l(s)$ as 
\begin{equation} 
\label{new_box_1} 
\Lambda_l(s)={1\over2\Pi(s)}\,{\partial\over\partial\xi_l}\Pi(s).   
\end{equation}  
Inserting $\Lambda_l(s)$ into the NI for vertices [Eq.~(\ref{res_4})] and 
solving for $\Delta_{l,f}^\mu(s)$, we find   
\begin{equation} 
\label{new_box_2} 
\Delta_{l,f}^\mu(s)={1\over\sqrt{\Pi(s)}}\,{\partial\over\partial\xi_l}\, 
{V_f^\mu(s)\over\sqrt{\Pi(s)}}, 
\end{equation} 
and analogously for $\Delta_{l,i}^\nu(s)$. 
Combining Eqs.~(\ref{box_1}) and (\ref{new_box_2}), the variation of 
$B_{fi}(s)$ with respect to the gauge parameter becomes 
\begin{equation} 
\label{new_box_3} 
{\partial\over\partial\xi_l}B_{fi}(s)
=i{\partial\over\partial\xi_l}\,{Q_{\mu\nu}V_f^\mu(s)V_i^\nu(s)\over\Pi(s)}.  
\end{equation} 
Eq.~(\ref{new_box_3}) implies that the most general gauge-independent 
combination is an arbitrary function of 
\begin{equation}
\label{box_7}
\Phi_{fi}(s)=-i{Q_{\mu\nu}V_f^\mu(s)V_i^\nu(s)\over\Pi(s)}+B_{fi}(s). 
\end{equation} 
Finally, requiring that the combination has a simple pole given by the zero of 
$\Pi(s)$, we see that it must be a linear function of $\Phi_{fi}(s)$.   
Thus, the NI tell us that, subject to the latter requirement, the most general 
combination that is gauge independent for arbitrary $s$ is the physical 
amplitude! 
This result can be readily extended to the case of field mixing. 

It is interesting to observe that the argument presented in this section may
be reversed to give a simple derivation of the functional structure of
Eqs.~(\ref{res_4}) and (\ref{box_1}).
Invoking the gauge independence of the $S$-matrix element given in
Eq.~(\ref{box_7}) and noting that $\partial B_{fi}(s)/\partial\xi_l$ does not
contain terms proportional to $1/\Pi(s)$, one concludes that the same is true
of
$-i\partial\left[Q_{\mu\nu}V_f^\mu(s)V_i^\nu(s)/\Pi(s)\right]/\partial\xi_l$.
Observing that $V_f^\mu(s)$ and $V_i^\nu(s)$ are independent functions and
employing Eq.~(\ref{ni_1}), one readily finds that
$\partial V_f^\mu(s)/\partial\xi_l$ satisfies a relation with the functional
structure of Eq.~(\ref{res_4}), subject to the constraint that the function
$\Delta_{l,f}^\mu(s)$ does not contain $1/\Pi(s)$ contributions.
Differentiating Eq.~(\ref{box_7}) with respect to $\xi_l$, inserting
Eqs.~(\ref{ni_1}) and (\ref{res_4}), and invoking once more the gauge
independence of the $S$ matrix, one obtains Eq.~(\ref{box_1}). 

\section{ANALYTICAL PROPERTIES AND NI}
  
As is well known, the analytical properties of the Green function $\Pi(s)$ 
permit the use of Cauchy's theorem, 
\begin{equation} 
\label{an_2} 
\Pi(s)={1\over2\pi i}\oint_\gamma ds^\prime{\Pi(s^\prime)\over s^\prime-s},   
\end{equation}  
where $\gamma$ is a closed contour in the complex $s^\prime$ plane that 
encircles the point $s$ counterclockwise. 
Eq.~(\ref{an_2}) and the distribution-based relation 
\begin{equation} 
\label{an_3} 
\lim_{\epsilon\to0}{1\over s-s^\prime+i\epsilon} 
=P{1\over s-s^\prime}-i\pi\delta(s-s^\prime),   
\end{equation} 
where $P$ denotes the principal value, leads to the derivation of dispersion 
relations and sum rules. 
 
Using Eq.~(\ref{an_2}) and the NI of Eq.~(\ref{ni_1}), we have 
\begin{eqnarray} 
\label{an_4} 
{\partial\over\partial\xi_l}\Pi(s)={1\over2\pi i}\oint_\gamma ds^\prime 
{2\Lambda_l(s^\prime)\Pi(s^\prime)\over s^\prime-s} 
={1\over(2\pi i)^2}\oint_\gamma ds^\prime\oint_{\gamma^\prime} 
ds^{\prime\prime}{2\Lambda_l(s^\prime)\Pi(s^{\prime\prime})\over(s^\prime-s) 
(s^{\prime\prime}-s^\prime)}, 
\end{eqnarray} 
where the contour $\gamma^\prime$ encircles $\gamma$ counterclockwise. 
Rewriting the factor $1/[(s^\prime-s)(s^{\prime\prime}-s^\prime)]$, the last
member of Eq.~(\ref{an_4}) becomes 
\begin{equation} 
\label{an_5} 
{\partial\over\partial\xi_l}\Pi(s)={1\over(2\pi i)^2}\oint_{\gamma} 
ds^\prime\oint_{\gamma^\prime}ds^{\prime\prime}2\Lambda_l(s^\prime) 
\Pi(s^{\prime\prime})\left[{1\over(s^\prime-s)(s^{\prime\prime}-s)} 
+{1\over(s^{\prime\prime}-s^\prime)(s^{\prime\prime}-s)}\right], 
\end{equation} 
where the second term vanishes, since the integration over $s^\prime$ along 
the contour $\gamma$ does not encircle any singularity. 
Therefore, we find 
\begin{equation} 
\label{an_6} 
{\partial\over\partial\xi_l}\Pi(s)=\left[{1\over2\pi i} 
\oint_{\gamma}ds^\prime{2\Lambda_l(s^\prime)\over s^\prime-s}\right]\Pi(s),  
\end{equation} 
which implies that $\Lambda_l(s)$ admits a spectral representation, analogous 
to Eq.~(\ref{an_2}), that is compatible with the analyticity of Green 
functions. 
  
A clear example of the analyticity of the functions $\Lambda_l(s)$ can be read 
off from the absorptive part of the Higgs-boson two-point function 
$\im\Pi_{HH}(s)$ presented in Ref.~\cite{si-kniehl}. 
From the imaginary part of Eq.~(\ref{ni_1}) at one loop,\footnote{%
The decoupling of unphysical modes, the mixing with the neutral would-be 
Goldstone boson, and the NI for Higgs bosons are discussed in Ref.~\cite{gg}.} 
we have 
\begin{equation} 
\label{higg_1} 
{\partial\over\partial\xi_W}\im\Pi^{(1)}_{HH}(s) 
=2\left(s-M_H^2\right)\im\Lambda_{W,H}^{H(1)}(s).   
\end{equation}  
Comparison with Ref.~\cite{si-kniehl} shows that 
\begin{eqnarray} 
\label{higg_1.1} 
\im\Lambda_{W,H}^{H(1)}(s)&=&\frac{GM_W^2}{2}\left(s+M_H^2\right)
\left[\frac{1}{2s} 
\left(1-\frac{4\xi_WM_W^2}{s}\right)^{-1/2}\theta\left(s-4\xi_WM_W^2\right) 
\right.\nonumber\\ 
&&{}-\left.\left(1-\frac{4\xi_WM_W^2}{s}\right)^{1/2} 
\delta\left(s-4\xi_WM_W^2\right)\right], 
\end{eqnarray} 
where $G=G_\mu/\left(2\pi\sqrt2\right)$, with $G_\mu$ being the muon decay 
constant.   
Although the second term seems to violate the analyticity of Green functions 
because of the presence the $\delta$ function, the factor 
$\left(1-4\xi_WM_W^2/s\right)^{1/2}$ leads to the vanishing of this 
contribution. 
Thus, we obtain 
\begin{equation} 
\label{higg_2} 
\im\Lambda_{W,H}^{H(1)}(s)=\frac{GM_W^2}{4}\left(1+\frac{M_H^2}{s}\right) 
\left(1-\frac{4\xi_WM_W^2}{s}\right)^{-1/2}\theta\left(s-4\xi_WM_W^2\right). 
\end{equation} 
The function $\Lambda_{W,H}^{H(1)}(s)$ can also be obtained by a direct 
evaluation of the relevant Feynman diagrams. 
In fact, Eq.~(\ref{higg_2}) has the structure one expects from such a  
computation. 
It shows that the one-loop Green function $\Lambda_{W,H}^{H(1)}(s)$ contains 
absorptive parts, but they are proportional to $\theta$ functions centered at   
some unphysical thresholds. 
This is due to the fact that, in the diagrams for $\Lambda_{W,H}^{H(1)}(s)$, 
only unphysical modes propagate at one loop. 
However, $\theta$ functions centered at physical thresholds may appear at 
higher orders. 
  
The NI tell us that the gauge dependence of a Green function is described by 
another Green function, which can be computed in terms of Feynman rules. 
The factorization implied by the NI is far from being trivial, and the  
Higgs-boson example shows how analyticity works in the factorization of the 
second member of the NI.   
  
\section{CONCLUSIONS}  
  
In this paper, we examined fundamental properties of unstable particles, such 
as their masses, widths, and partial widths, in the light of the NI, which 
describe the gauge dependence of Green functions.    
  
In Sec.~II, we applied the NI to show that the conventional 
definitions of mass and width of unstable particles are gauge dependent in 
NNLO. 
This shows that, in the gauge-theory context, the conventional treatment of 
unstable particles is strictly valid through NLO. 
For completeness, in the same section, we revisited the formal proof, to all  
orders, of the gauge independence of the pole position $\bar s$ \cite{gg}. 
 
In Sec.~III, we applied the NI for vertex functions to prove the 
gauge independence of the pole residues. 
This motivates a gauge-independent definition of partial widths. 
As explained in Ref.~\cite{Grassi:2001dz}, this definition does not satisfy 
the additivity property in NNLO. 
However, this problem can be circumvented by a judicious rescaling of the 
partial widths.  
 
We then considered an alternative definition of branching ratios $\tilde{B}_f$ 
and partial widths $\tilde{\Gamma}_f$ that are manifestly additive and closely 
resemble the conventional ones. 
Using the NI, we showed that $\tilde{B}_f$ and $\tilde{\Gamma}_f$ are gauge 
independent through NNLO. 
This explains a significant result obtained in Ref.~\cite{Grassi:2001dz}, 
namely that the cross section of $e^+e^-$ annihilation at the $Z^0$-boson peak 
in the SM, expressed in terms of $\tilde{\Gamma}_f$, is gauge independent 
through NNLO. 
Although this result is sufficient for the phenomenological requirements of 
electroweak physics in the foreseeable future, $\tilde{B}_f$ and 
$\tilde{\Gamma}_f$ are not expected to be gauge independent in still higher 
orders.  
  
We also used the NI to show that the usual assumption that 
$\im A\left(m_2^2\right)$ can be expressed as a sum of physical cut 
contributions fails in NNLO. 
The difference between these two quantities is given by a function  
$G\left(m_2^2\right)$ that emerges in NNLO. 
In Ref.~\cite{Grassi:2001dz}, $G\left(m_2^2\right)$ was shown to be 
non-vanishing by studying the difference between two gauge-independent 
definitions of total width based, respectively, on the pole residues and the 
pole position. 
In the present paper, we employed the NI to derive an expression for 
$G\left(m_2^2\right)$ with the same mathematical structure.  
 
In Sec.~IV, we showed how to modify the alternative definition 
of branching ratios, discussed in Sec.~III, in order to extend 
the gauge independence to all orders. 
 
In Sec.~V, we discussed the NI for box diagrams and showed that 
the physical amplitude is the most general combination of self-energy, vertex, 
and box contributions that is gauge independent for arbitrary $s$!
Reversing the argument, we also showed that the functional structure of the NI
for vertex and box functions can be derived starting from the gauge
independence of the $S$ matrix and well-known properties of the box
amplitudes.
It should also be emphasized that the gauge independence of the physical 
amplitude implies the same property for the coefficients of its Laurent 
expansion. 
 
Section~VI discusses the analytic properties of the Green functions 
$\Lambda_l(s)$ that play an important role in the NI. 
 
The Appendix gives the explicit one-loop relations between the functions  
$\Lambda_{l,\alpha}^\delta(s)$ and $\Delta_{l,f}^{\delta,\mu}(s)$ that occur 
in the NI and the calculations of Ref.~\cite{deg} in the $Z$-$\gamma$ sector 
of the SM. 

\vspace{1cm}
\begin{center}
{\bf Acknowledgements}
\end{center}
\smallskip

B.A.K. and A.S. thank the members of the Physics Department of the University  
of Barcelona and the Theoretical Physics Division of CERN, respectively, for 
their kind hospitality during September 2001. 
The research of P.A.G. and A.S. was supported in part by National Science 
Foundation through Grant No.~PHY-0070787. 
The research of B.A.K. was supported in part by the Deutsche 
Forschungsgemeinschaft through Grant No.\ KN~365/1-1, by the 
Bundesministerium f\"ur Bildung und Forschung through Grant No.\ 05~HT1GUA/4, 
and by the European Commission through the Research Training Network 
{\it Quantum Chromodynamics and the Deep Structure of Elementary Particles}. 

\renewcommand {\theequation}{\Alph{section}.\arabic{equation}}
\begin{appendix}

\setcounter{equation}{0}
\section{ONE-LOOP CONTRIBUTIONS TO NI FUNCTIONS}
\label{appA}  
 
In this Appendix, we present the one-loop contributions to the NI for the 
$Z$-$\gamma$ sector of the SM. 
In particular, we discuss the relation between the NI Green functions 
$\Lambda_{l,\alpha}^\delta(s)$ and $\Delta_{l,f}^{\delta,\mu}(s)$ defined in 
Eqs.~(\ref{ni_2}) and (\ref{res_5}), respectively, and the one-loop 
computations of Ref.~\cite{deg}.   
 
According to Eq.~(\ref{ni_2}), in the $Z$-$\gamma$ case, the transverse 
self-energies $\Pi_{ZZ}(s)$, $\Pi_{Z\gamma}(s)$, and $\Pi_{\gamma\gamma}(s)$ 
satisfy 
\begin{eqnarray} 
\label{app_1} 
{\partial\over\partial\xi_l}\Pi_{ZZ}(s)&=& 
2\sum_{\delta=Z,\gamma}\Lambda_{l,Z}^\delta(s)\Pi_{\delta Z}(s), 
\nonumber\\  
{\partial\over\partial\xi_l}\Pi_{Z \gamma}(s)&=& 
\sum_{\delta=Z,\gamma}\left[\Lambda_{l,Z}^\delta(s)\Pi_{\delta\gamma}(s) 
+\Lambda_{l,\gamma}^\delta(s)\Pi_{\delta Z}(s)\right], 
\nonumber\\  
{\partial\over\partial\xi_l}\Pi_{\gamma\gamma}(s)&=& 
2\sum_{\delta=Z,\gamma}\Lambda_{l,\gamma}^\delta(s)\Pi_{\delta\gamma}(s), 
\end{eqnarray}  
where $\Lambda_{l,Z}^Z(s)$, $\Lambda_{l,Z}^\gamma(s)$, 
$\Lambda_{l,\gamma}^Z(s)$, and $\Lambda_{l,\gamma}^\gamma(s)$ involve the 
gauge fields $Z$ or $\gamma$ (lower index), the BRST transformation of the 
gauge parameter $\xi_l$, and the sources of the BRST variations associated
with $Z$ or $\gamma$ (upper index). 
As implied by the statement before Eq.~(\ref{pole_mixed}), the complex pole 
positions $\bar s$ are given by the equation 
\begin{equation} 
\label{app_1.1} 
\Pi_{ZZ}(\bar s)\Pi_{\gamma\gamma}(\bar s)-\Pi_{Z\gamma}^2(\bar s)=0, 
\end{equation} 
which, being a quadratic polynomial in $\bar s$ at tree level, has two 
solutions at all orders. 
Actually, one solution is trivially $\bar s=0$, due to BRST symmetry 
(cf.\ Ref.~\cite{brs}). 
 
In order to compare this with the explicit one-loop computations of 
Ref.~\cite{deg}, we reduce Eq.~(\ref{app_1}) to the one-loop level using the 
fact that the $\Lambda_{l,\alpha}^\delta(s)$ and $\Pi_{Z\gamma}(s)$ functions 
are of $O(g^2)$: 
\begin{eqnarray} 
\label{app_2} 
{\partial\over\partial\xi_l}\Pi_{ZZ}^{(1)}(s) 
&=&2\left(s-M_Z^2\right)\Lambda_{l,Z}^{Z(1)}(s), 
\nonumber\\ 
{\partial\over\partial\xi_l}\Pi_{Z\gamma}^{(1)}(s) 
&=&\left[s\Lambda_{l,Z}^{\gamma(1)}(s) 
+\left(s-M_Z^2\right)\Lambda_{l,\gamma}^{Z(1)}(s)\right], 
\nonumber\\ 
{\partial\over\partial\xi_l}\Pi_{\gamma\gamma}^{(1)}(s) 
&=&2s\Lambda_{l,\gamma}^{\gamma(1)}(s). 
\end{eqnarray}  
Comparing Eq.~(\ref{app_2}) with the results of Ref.~\cite{deg} and employing 
the functions $v_W(s)$ and $\eta_W(s)$ defined in that work, we obtain 
\begin{eqnarray} 
\label{app_3} 
\int_1^{\xi_W}d\xi_W^\prime\Lambda_{W,Z}^{Z(1)}\left(s,\xi_W^\prime\right) 
&=&-g^2c_w^2(\xi_W-1)\left[v_W(s)+
\frac{1}{2}\left(s-M_Z^2\right)\eta_W(s)\right], 
\nonumber\\ 
\int_1^{\xi_W}d\xi_W^\prime\Lambda_{W,Z}^{\gamma(1)}\left(s,\xi_W^\prime\right) 
&=&-g^2s_wc_w(\xi_W-1)\left[v_W(s)
+\frac{1}{2}\left(s-M_Z^2\right)\eta_W(s)\right], 
\nonumber\\ 
\int_1^{\xi_W}d\xi_W^\prime\Lambda_{W,\gamma}^{Z(1)}\left(s,\xi_W^\prime\right) 
&=&-g^2s_wc_w(\xi_W-1)\left[v_W(s)+\frac{s}{2}\eta_W(s)\right], 
\nonumber\\ 
\int_1^{\xi_W}d\xi_W^\prime\Lambda_{W,\gamma}^{\gamma(1)} 
\left(s,\xi_W^\prime\right) 
&=&-g^2s_w^2(\xi_W-1)\left[v_W(s)+\frac{s}{2}\eta_W(s)\right],
\end{eqnarray}
where $s_w\equiv\sin\theta_w$.
The other functions $\Lambda_{l,\alpha}^{\delta(1)}(s,\xi_k)$, with $l\neq W$
and $\alpha,\delta=Z,\gamma$, vanish. 
From Eq.~(\ref{app_3}), we can immediately see that there are only two 
functionally independent Green functions, viz  
\begin{eqnarray} 
\label{app_4} 
\Lambda_{W,Z}^{3(1)}(s,\xi_W)&=&\Lambda_{W,Z}^{Z(1)}(s,\xi_W) 
+{s_w\over c_w}\Lambda_{W,Z}^{\gamma(1)}(s,\xi_W), 
\nonumber\\ 
\Lambda_{W,\gamma}^{3(1)}(s,\xi_W)&=&{c_w\over s_w}   
\Lambda_{W,\gamma}^{Z(1)}(s,\xi_W)+\Lambda_{W,\gamma}^{\gamma(1)}(s,\xi_W). 
\end{eqnarray}  
This result is expected on the basis of BRST symmetry. 
In fact, following Ref.~\cite{brs}, one only introduces the source coupled to 
non-linear BRST transformations; for example, one needs the source for the 
BRST variations of the gauge bosons $W^i_\mu$ of the $SU(2)$ triplet, but not 
for the abelian gauge field $B_\mu$. 
This implies that there are only two independent functions 
$\Lambda_{l,\alpha}^\delta(s,\xi_k)$ in the $Z$-$\gamma$ sector. 
 
As for the vertex functions, from Eq.~(\ref{res_5}) at one loop, we have 
\begin{eqnarray} 
\label{app_5} 
{\partial\over\partial\xi_l}\Gamma_{Z,f}^{\mu(1)}(s) 
&=&\sum_{\delta=Z,\gamma}\Lambda_{l,Z}^{\delta(1)}(s)
\Gamma_{\delta,f}^{\mu(0)}(s) 
+\left(s-M_Z^2\right)\Delta_{l,f}^{Z,\mu(1)}(s), 
\nonumber\\ 
{\partial\over\partial\xi_l}\Gamma_{\gamma,f}^{\mu(1)}(s) 
&=&\sum_{\delta=Z,\gamma}\Lambda_{l,\gamma}^{\delta(1)}(s)
\Gamma_{\delta,f}^{\mu(0)}(s) 
+s\Delta_{l,f}^{\gamma,\mu(1)}(s). 
\end{eqnarray}  
Comparing with the results of Ref.~\cite{deg}, we find 
\begin{eqnarray} 
\label{app_6} 
\int_1^{\xi_W}d\xi_W^\prime\Delta_{W,f}^{Z,\mu(1)}\left(s,\xi_W^\prime\right) 
&=&{g^2c_w^2\over2}(\xi_W-1)\eta_W(s)\left[\Gamma_{Z,f}^{\mu(0)}(s) 
+{s_w\over c_w}\Gamma_{\gamma,f}^{\mu(0)}(s)\right], 
\nonumber\\ 
\int_1^{\xi_W}d\xi_W^\prime\Delta_{W,f}^{\gamma,\mu(1)} 
\left(s,\xi_W^\prime\right) 
&=&{g^2s_wc_w\over2}(\xi_W-1)\eta_W(s)\left[\Gamma_{Z,f}^{\mu(0)}(s) 
+{s_w\over c_w}\Gamma_{\gamma,f}^{\mu(0)}(s)\right]. 
\end{eqnarray}  
The other functions $\Delta_{l,f}^{\delta,\mu(1)}(s,\xi_k)$, with $l\neq W$ 
and $\delta=Z,\gamma$, vanish. 
We note again that $\Delta_{W,f}^{\delta,\mu(1)}(s,\xi_W)$, with 
$\delta=Z,\gamma$, are not independent, but proportional to  
\begin{equation} 
\label{app_7} 
\Delta_{W,f}^{3,\mu(1)}(s,\xi_W)=\Delta_{W,f}^{Z,\mu(1)}(s,\xi_W) 
+\frac{s_w}{c_w}\Delta_{W,f}^{\gamma,\mu(1)}(s,\xi_W), 
\end{equation} 
which is expected from BRST symmetry, as explained above. 
Moreover, we see that they are proportional to a combination of tree-level 
vertex functions. 
By inserting Eqs.~(\ref{app_3}) and (\ref{app_6}) in Eq.~(\ref{app_5}), we 
obtain 
\begin{eqnarray} 
\label{app_8} 
{\partial\over\partial\xi_W}\Gamma_{Z,f}^{\mu(1)}(s,\xi_W)&=& 
-g^2c_w^2{\partial\over\partial\xi_W}[(\xi_W-1)v_W(s)] 
\left[\Gamma_{Z,f}^{\mu(0)}(s)
+{s_w\over c_w}\Gamma_{\gamma,f}^{\mu(0)}(s)\right], 
\nonumber\\ 
{\partial\over\partial\xi_W}\Gamma_{\gamma,f}^{\mu(1)}(s,\xi_W)&=& 
-g^2s_wc_w{\partial\over\partial\xi_W}[(\xi_W-1)v_W(s)] 
\left[\Gamma_{Z,f}^{\mu(0)}(s) 
+{s_w\over c_w}\Gamma_{\gamma,f}^{\mu(0)}(s)\right], 
\end{eqnarray} 
in agreement with the vertex computations in Ref.~\cite{deg}. 

\end{appendix}


\begin{thebibliography}{99}

\bibitem{si_1} A. Sirlin, 
Phys.\ Rev.\ Lett.\ {\bf67}, 2127 (1991);
Phys.\ Lett.\ B {\bf267}, 240 (1991).  
    
\bibitem{zmass} M. Consoli and A. Sirlin, 
in {\it Physics at LEP}, CERN Yellow Report No.~86-02, 1986, Vol.~1, p.~63;  
S. Willenbrock and G. Valencia, 
Phys.\ Lett.\ B {\bf259}, 373 (1991); 
R. G. Stuart, 
Phys.\ Lett.\ B {\bf262}, 113 (1991); {\bf272}, 353 (1991);
Phys.\ Rev.\ Lett.\ {\bf70}, 3193 (1993);
Nucl.\ Phys.\ {\bf B498}, 28 (1997); 
H. Veltman, 
Z. Phys.\ C {\bf62}, 35 (1994);
A. R. Bohm and N. L. Harshman, 
Nucl.\ Phys.\ {\bf B581}, 581 (2000).  
  
\bibitem{psz} M. Passera and A. Sirlin, 
Phys.\ Rev.\ Lett.\ {\bf77}, 4146 (1996).  
  
\bibitem{si-kniehl} B. A. Kniehl and A. Sirlin, 
Phys.\ Rev.\ Lett.\ {\bf81}, 1373 (1998);
Phys.\ Lett.\ B {\bf440}, 136 (1998). 
  
\bibitem{gg} P. Gambino and P. A. Grassi, 
Phys.\ Rev.\ D {\bf62}, 076002 (2000). 
  
\bibitem{bhatta} B. A. Kniehl,
Nucl.\ Phys.\ {\bf B357}, 439 (1991);  
T. Bhattacharya and S. Willenbrock, 
Phys.\ Rev.\ D {\bf47}, 4022 (1993);
B. A. Kniehl, C. P. Palisoc, and A. Sirlin,  
Nucl.\ Phys.\ {\bf B591}, 269 (2000).  
  
\bibitem{ps} M. Passera and A. Sirlin, 
Phys.\ Rev.\ D {\bf58}, 113010 (1998).  
  
\bibitem{Grassi:2001dz} P. A. Grassi, B. A. Kniehl, and A. Sirlin,  
Phys.\ Rev.\ Lett.\ {\bf86}, 389 (2001).  

\bibitem{kon} W. Konetschny and W. Kummer,
Nucl.\ Phys.\ {\bf B100}, 106 (1975).

\bibitem{nie} N. K. Nielsen,
Nucl.\ Phys.\ {\bf B101}, 173 (1975).

\bibitem{klu} H. Kluberg-Stern and J. B. Zuber,  
Phys.\ Rev.\ D {\bf12}, 467 (1975).

\bibitem{pig} H. Kluberg-Stern and J. B. Zuber, 
Phys.\ Rev.\ D {\bf12}, 482 (1975);
O. Piguet and K. Sibold, 
Nucl.\ Phys.\ {\bf B253}, 517 (1985);
J. C. Breckenridge, M. J. Lavelle, and T. G. Steele,
Z.\ Phys.\ C {\bf65}, 155 (1995);
R. H\"au\ss ling, E. Kraus, and K. Sibold,
Nucl.\ Phys.\ {\bf B539}, 691 (1999);
O. M. Del Cima, D. H. T. Franco, and O. Piguet,
Nucl.\ Phys.\ {\bf B551}, 813 (1999).

\bibitem{kum} W. Kummer,
Eur.\ Phys.\ J. C {\bf21}, 175 (2001).

\bibitem{st} J. C. Taylor,
Nucl.\ Phys.\ {\bf B33}, 436 (1971);
A. A. Slavnov,
Teor.\ Mat.\ Fiz.\ {\bf10}, 153 (1972)
[Theor.\ Math.\ Phys.\ {\bf10}, 99 (1972)].
  
\bibitem{si_80} A. Sirlin, 
Phys.\ Rev.\ D {\bf22}, 971 (1980).  
  
\bibitem{deg} G. Degrassi and A. Sirlin,  
Nucl.\ Phys.\ {\bf B383}, 73 (1992).  
   
\bibitem{brs} K.-I. Aoki, Z. Hioki, R. Kawabe, M. Konuma, and T. Muta, 
Prog.\ Theor.\ Phys.\ Suppl.\ {\bf73}, 1 (1982);
E. Kraus, 
Ann.\ Phys.\ (N.Y.) {\bf262}, 155 (1998);
P. A. Grassi,
Nucl.\ Phys.\ {\bf B560}, 499 (1999). 
  
\end{thebibliography}
\end{document}